\documentclass[%
reprint,
 showpacs,showkeys,
 amsmath,amssymb,
 aps,
]{revtex4-1}
\usepackage{amsmath}
\usepackage{cases}
\usepackage{amssymb}
\usepackage{CJK}
\usepackage{graphicx}
\usepackage{dcolumn}
\usepackage{bm}
\usepackage{color}
\usepackage[pagewise]{lineno}
\usepackage{subfigure}

\begin{document}
\begin{CJK*}{GBK}{} 

\preprint{APS/123-QED}

\title{Bayesian evaluation of residual production cross sections in proton induced spallation reactions
}
\author{Dan Peng$^{1}$}
\author{Hui-Ling Wei$^{1}$}
\author{Xi-Xi Chen$^{1}$}
\author{Xiao-Bao Wei$^{1}$}
\author{Yu-Ting Wang$^{1,2}$}
\author{Jie Pu$^{1,2}$}
\author{Kai-Xuan Cheng$^{1,2}$}
\author{Chun-Wang Ma$^{1,2}$}
\thanks{Corresponding author. Email address: machunwang@126.com}

\affiliation{
$^{1}$ School of Physics, Henan Normal University, Xinxiang \textit{453007}, China\\
$^{2}$ Institute of Particle and Nuclear Physics, Henan Normal University, Xinxiang \textit{453007}, China\\
}

\date{\today}

\begin{abstract}

\noindent\textbf{Background}: Residual production cross sections of spallation reactions are key infrastructure data for nuclear physics and related applications. Due to the complexity of the reaction mechanism, a wide range of incident energies and a abundance of fragments involved in the reactions, it is a challenge to obtain accurate and complete energy-dependent residual cross sections.

\noindent\textbf{Purpose}: To propose a physical guided machine learning model based on Bayesian neural network (BNN), which can be used to predict the excitation curve for fragments, residual cross sections in proton induced spallation reactions.

\noindent\textbf{Method}: A simplified version of EPAX empirical parameterizations (sEPAX) has been introduced as the physical guidance for the BNN learning. Two types of sample data for measured proton-induced residual production cross sections have been adopted, i.e., (1) the fragment excitation functions for reactions up to 2.6 GeV/u, and (2) the isotopic cross sections for reactions below 1 GeV/u. Both the BNN and ``BNN + sEPAX'' methods have been used to construct the predictive models for the proton-induced spallation reactions.

\noindent\textbf{Result}: The isotopic distributions and mass cross sections have been compared for 365 MeV/u $^{40}$Ca + $p$ and 1 GeV/u $^{136}$Xe + $p$ reactions, and excitation functions for $^{22}$Na, $^{24}$Na, $^{28}$Mg, $^{26}$Al in $^{40}$Ca + $p$ reaction, $^{24}$Na, $^{36}$Cl, $^{47}$Sc, and $^{52}$Mn in $^{56}$Fe + $p$ reaction, $^{54}$Mn, $^{75}$Se, $^{105}$Ag, $^{138}$Ba in $^{138}$Ba + $p$ reaction and $^{85}$Sr, $^{95}$Nb, $^{160}$Er, $^{173}$Hf in $^{197}$Ag + $p$ reaction. It is found that BNN method needs sufficient information to achieve good extrapolation, while BNN + sEPAX method can perform better extrapolation based on less information due to the physical guidance of sEPAX formula.

\noindent\textbf{Conclusions}: The BNN + sEPAX method can reasonably extrapolate with less information compared with BNN method. The BNN + sEPAX method provides a new approach to predict the energy-dependent residual cross sections produced in proton-induced spallation reactions from tens of MeV/u up to several GeV/u.

\end{abstract}

\pacs{25.70.Pq, 25.70.Mn, 21.65.Cd}

\keywords{BNN, EPAX, spallation reaction, intermediate mass fragments}

\maketitle
\end{CJK*}


\section{introduction}

The residual fragments cross sections in spallation reactions are key infrastructure data for nuclear applications in many aspects, such as
nuclear physics, radiation damage to electronics and radio-protection of astronauts \cite{r.protect}, extraterrestrial bodies history via the radioisotopes produced inside \cite{iso.history}, tracing the transport history of cosmic rays \cite{ray.history} and abundance of Li, Be, and B elements \cite{taleofnuclei}, neutron sources or as radioactive isotope beams like the China Spallation Neutron Source (CSNS) facility \cite{CSNS}, the Beijing Rare Ion beam Facility (BRIF) facility \cite{BRIF}, the Accelerator-Driven System (ADS) \cite{ADS1999,ADS2000}, and even in situ proton therapy tomography \cite{PT1,PT2}. Traditional methods to predict fragment productions in spallation reactions include the transport models like quantum molecular dynamics (QMD) \cite{QMD2009,QMD2020,QMD2019}, statistical muti-fragmentation model (SMM) \cite{SMM1995,SMM2001,SMM2005} and the Li\`{e}ge intranuclear cascade (INC) \cite{INCL2013,INCL2014,INCL2015} model, etc. A de-excitation (principally evaporation or fission) simulation is always performed after the QMD, SMM, INC models for better predictions. Semi-empirical formula like EPAX \cite{EPAX1,EPAX2,EPAX3} and SPACS \cite{SPACS} have also been frequently adopted to predict the fragment cross sections in spallation reactions for incident energy higher than 100 MeV/u. Most of the models mentioned above have participated in the international benchmark done under the OECN/NEA \cite{NEA1993,NEA1994,NEA1995,NEA1997} in the mid-nineties and the auspices of the International Atomic Energy Agency (IAEA) in 2010. However, difficulties still exist for the reasons that a wide range of incident energy and a vast of fragments are involved in the spallation reactions. The inadequate precise of present models prevent their applications in many key problems.

Machine learning is efficient to form new model based on the big-data learning, which has been involved in various industries and basic science researches, such as data mining, medical diagnosis, handwriting recognition, biological field, engineering application, automatic driving, stock analysis, and so on. Machine learning techniques have made various novel applications in physics, e.g., nuclear mass \cite{Nmass1,Nmass2,Nmass3,Nmass4}, nuclear charge radii \cite{Nradii}, nuclear $\beta$-decay half-life \cite{Nhalf-life}, neutron-induced reactions \cite{N-induced}, fission yields \cite{Nfission,QiaoChY}, emitting nuclei \cite{N-emitting}, quantum many-body problem \cite{QMB2017,QMB2020}, strong gravitational lenses \cite{SGL}, phases of matter \cite{PM1,PM2,PM3,PM4}, temperature determination in heavy-ion collision \cite{YDSong}, single crystal growth \cite{SCG}, experimental control \cite{EXPC} and nuclear liquid-gas phase transition \cite{RWang2020}. As one of the machine learning techniques, the Bayesian neural networks (BNN) have many advantages, such as automatic complexity control, possibility to use prior information and hierarchical models for the hyperparameters, predictions for outputs and giving uncertainty qualification \cite{Neal1,Neal2}.

Effort has also been paid for the construction of new approaches to describe spallation reactions using the BNN technologies. The direct learning by BNN and the physical guided BNN + SPACS approaches have been performed in previous works \cite{Nspallation1,Nspallation2}. In them, the adopted database has about 4,000 data, which have been measured at GSI, Darmstadt \cite{GSI-Fe,GSI-Xe200,GSI-Xe500,GSI-Xe1000}, and the Lawrence Berkeley Laboratory (LBL) \cite{LBL-Ar,LBL-Ca}. Because the deficient information about incident energies and light fragments in the sample data, unphysical predictions have been found in BNN method. With the guidance of SPACS empirical formula, good predictions have been found in the BNN + sEPAX method for reactions within the incident energies ranging from 300 MeV/u to 1 GeV/u, but poor extrapolation results arise for light fragments and incident energies lower than 300 MeV/u. 

In this work, to improve the model accuracy and generalization ability, more than 10,000 cross sections measured at RIKEN \cite{RIKEN-Nb113,RIKEN-Zr105,RIKEN-Pd118+196,RIKEN-Cs+Sr} and excitation functions for proton induced reactions \cite{Michel2014} have been incorporated. To provided reasonable physical guides, a simplified EPAX formula (named as sEPAX) will be proposed. A physical guided BNN + sEPAX model will be constructed to quantify the patterns of systematic deviations between theory and experiment 
The paper is organized as follows. In Section \ref{method}, the Bayesian theory and sEPAX formula will be briefly introduced. In Section \ref{result}, the accuracy and generalization of BNN and BNN + sEPAX models are demonstrated. Finally, conclusion are presented in Section \ref{summary}.

\section{model descriptions}
\label{method}
The main concepts of the Bayesian method, the simplified EPAX formula and the model structure for BNN and BNN + sEPAX method will be briefly introduced in this section.
\subsection{Bayesian method\label{bnn}}
The BNN method is one of the typical multilayer perceptron (MLP) networks. It is a ``back-propagation'' or ``feedforward'' network. In the general ideas of MLP networks, some numbers of layers of hidden unities will be constructed, via which the output values $f_{k}(x)$ can be exported from the input sets $x^{(i)}$. In a typical network of one hidden layer, the outputs are calculated as follows, 
\begin{equation}\label{MLP}
f_{k}(x;\theta)= a_{k}+\sum_{j=1}^{H}b_{jk} \tanh(c_j+\sum_{i=1}^{I} d_{ji} x_i),
\end{equation}
where $H$ denotes the number of hidden unites, and $I$ is the number of input variables $x=\{x^{(i)}\}$. $d_{ij}$($b_{jk}$) are the weights on the connection from input unit $i$ (hidden unite $j$) to hidden unit $j$ (output unites $k$). The $c_{j}$ and $a_{k}$ are the biases of the hidden and output unites. The weights and bias are the parameters of the network, i.e., $\theta =\{a, b_{j}, c_j, d_{ji}\}$. Each output $f_{k}(x)$ is a weighted sum of hidden unit values plus a bias. Each hidden unite computes a similar weight sum of input values, and then passes it through a nonlinear activation function. The activation function has chosen to be the hyperbolic tangent ($\tanh$) in this work.

For a regression task involving the prediction of a noisy vector $t$ of target variables given the input vector $x$, the likelihood function $p(D/\theta)$ might be defined to be Gaussian-type, with $t$ having a mean of $f_{k}(x)$ (k=1) and a standard deviation of $\alpha$,
\begin{equation}\label{likelihood}
\begin{split}
 p(D|\theta)=\exp(-\chi^{2}/2),   \\
\chi^{2}=\sum\limits_{i}^{N} [t^{(i)}-f(x^{(i)};\theta)]^2 / {\alpha_{i}^2},
\end{split}
\end{equation}
where $D=(x^{(i)},t^{(i)})$ ($i =$ 1, 2, ... , $N$), and $x^{(i)}$ ( $t^{(i)}$) are the inputs (outputs) of the network structure. 

\subsection{Simplified EPAX formula (sEPAX)}
\label{sEPAX}
Following the EPAX parametrizations in Ref. \cite{EPAX1}, the cross section ($\sigma$) for a fragment with mass and charge ($A, Z$) produced from a projectile nucleus ($A_{p}$, $Z_{p}$) impinging on a target nucleus (with $A_{t}=Z_{t}=1$ for proton) is written as,
\begin{equation}
\sigma(A,Z) = Y_{A}\sigma_{Z}(Z_{prob}-Z) = Y_{A}n \exp(-R\mid Z_{prob}-Z\mid^{U}), \label{SigAZ}
\end{equation}
in which $Y_{A}$ is the mass yield. $\sigma_{Z}$ is the ``charge dispersion'' referring to the elemental distribution of given mass number $A$ around the maximum of charge dispersion $Z_{prob}$. The shape of the charge dispersion is governed by the width parameter $R$ and the exponent $U$. The normalization factor $n=\sqrt{R/\pi}$ assures the unity of the integral charge dispersion.

The mass yield, $Y_{A}$, is assumed to exponentially depend on the mass difference between projectile and fragment ($A_{p}-A$),
\begin{equation}
Y_A = S\cdot P\cdot \exp[-P\cdot(A_{p}-A)].
\end{equation}
The slope $P$ depends on the mass of the projectile,
\begin{equation}
P=\exp(-1.731-0.01339\cdot A_p).
\end{equation}
An overall scaling factor $S$ accounts for the peripheral reaction, which depends both on the mass of projectile and target nuclei,
\begin{equation}
S=0.27[(A_p)^{1/3} + (A_t)^{1/3}-1.8].
\end{equation}

The parameters $R$, $Z_{prob}$, and $U$ are strongly correlated to each other, which are difficult to be uniquely obtained using the least-squares fitting technique. $U$ is assorted to $U_{p}$ and $U_{n}$ for the neutron-deficient and neutron-rich sides of the valley of $\beta$-stability, which have different values in the three EPAX versions \cite{EPAX1,EPAX2,EPAX3}. In this work, $U=U_{p}=U_{n}=2.0$ are taken for simplification. $Z_{prob}$ is parameterized according to the $\beta$-stability line,
\begin{equation}
Z_{prob} \simeq Z_\beta = A/(1.98+0.0155\cdot A^{2/3}).
\end{equation}
Similar to $Z_{prob}$, the width parameter $R$ is taken to depend on the mass of fragment,
\begin{equation}
R \simeq \exp(-0.015\cdot A+3.2\times 10^{5}\cdot A^2).
\end{equation}

\subsection{Model construction}
\label{construction}

Two types of database for spallation reactions have been adopted in this work.  One is the residual production cross sections from various spallation reactions measured by bombarding one projectile at hundreds of MeV/u on a liquid-hydrogen target using the reverse kinematics technique at GSI, LBL and RIKEN (as listed in Table \ref{measdata}), which are named as D1. The other one is the data of excitation functions for fragments produced in proton induced reactions aiming at describing productions of cosmogenic nuclides in extraterrestrial matter by solar and galactic cosmic ray protons, which are named as D2. The D2 cover productions of nuclides from $^{nat}$C, $^{nat}$N, $^{nat}$O $^{nat}$F, $^{nat}$Mg,$^{27}$Al, $^{nat}$Si, $^{nat}$Ca, $^{nat}$Ti, $^{nat}$V, $^{55}$Mn, $^{nat}$Fe, $^{59}$Co, $^{nat}$Ni $^{nat}$Cu, $^{nat}$Sr, $^{89}$Y, $^{nat}$Zr, $^{93}$Nb, $^{nat}$B, and $^{197}$Au \cite{Michel2014}. In D1 the number of data is about 10,000, while in D2 it is about 3,000. It should be noted that only the reactions of incident energy higher than 30 MeV/u and the measuring uncertainty less than 30 percent are adopted in the learning set. Besides, the data for pick-up fragments are excluded from the learning set.

\begin{table}[thbp]
\caption{(Color online) A list of the adopted data for the measured residue cross sections in the $X$ + p spallation reactions at GSI, LBL and RIKEN.}
\label{measdata}
\centering
\begin{tabular}{p{60pt}<{\centering}|p{60pt}<{\centering}|p{60pt}<{\centering}|p{60pt}<{\centering}}
  \hline
  \hline
    $^{A}X$ + $p$  &E(MeV/u)  & Charge Range &Reference\\
  \hline
                     &300; 500; 750;  &    \\
  $^{56}$Fe + $p$    &1000; 1500 &8-27    &\cite{GSI-Fe} \\
  \hline
  $^{36}$Ar + $p$  &361; 546; 765  &9-17     &\cite{LBL-Ar}  \\

  \hline
  $^{40}$Ar + $p$  &352    &9-17       &\cite{LBL-Ar} \\
  \hline
  $^{40}$Ca + $p$   &356; 565; 763  &10-20   &\cite{LBL-Ca} \\

  \hline

                     &200  &48-55 &\cite{GSI-Xe200} \\
  $^{136}$Xe + $p$   &500  &41-56 &\cite{GSI-Xe500} \\
                     &1000 &3-56 &\cite{GSI-Xe1000} \\
 \hline

  $^{113}$Nb + $p$  &113 &37-42  &\cite{RIKEN-Nb113} \\

  \hline

  $^{93}$Zr + $p$  &105  &36-41 &\cite{RIKEN-Zr105}  \\

  \hline
  $^{107}$Pd + $p$  &118; 196 &42-47  &\cite{RIKEN-Pd118+196}\\
  \hline
  $^{137}$Cs + $p$   &185 &51-56  &\cite{RIKEN-Cs+Sr}   \\
  $^{90}$Sr + $p$  &185  &34-39\\
  \hline
  \hline
\end{tabular}\\
\end{table}

In constructing the models, the minimum numbers of parameters in the input set is chosen, which are $\{x^{(i)}\} = \{A_{p}^{(i)},Z_{p}^{(i)},E^{(i)},Z_{f}^{(i)},A_{f}^{(i)}$\}, with $Z_{p(f)}^{(i)}$($A_{p(f)}^{(i)}$) being the mass (charge) number of the projectile (fragment), and $E^{(i)}$ the incident energies. The output set is $t^{(i)}$ = lg(${\sigma}_{exp}^{(i)}$) for BNN method, and $t^{(i)}$ = lg(${\sigma}_{exp}^{(i)}$)- lg($\sigma_{th}^{(i)}$) for BNN + sEPAX method, with $\sigma^{(i)}_{exp}$ being the measured data and $\sigma^{(i)}_{th}$ the theoretical calculations by sEPAX formula. A 5-32-1 structure is adopted both for BNN and BNN + sEPAX models, i.e., 5 inputs $\{x^{(i)}\} = \{A_{p}^{(i)},Z_{p}^{(i)},E^{(i)},Z_{f}^{(i)},A_{f}^{(i)}$\}, one single layer with 32 hidden unites and one output set $t^{(i)}$ = lg(${\sigma}_{exp}^{(i)}$). 
It is noted that the incident energy is the main variable in the D2 dataset, which accounts for the main weight in the network, and the weight of fragments is very small. In the D1 dataset, the weights are contrary to that of D2. Since the data in D1 and D2 have significant difference, in  constructing the BNN and BNN + sEPAX models, the D2 dataset ($N_{l}$ = 11,807) is firstly adopted as the learning set and D1 as the testing set to verify whether the have the same fragment production mechanisms.

\section{Results and discussion}
\label{result}
Based on the constructed BNN and BNN + sEPAX models, the discussion will be concentrated on the isotopic distributions, mass distributions, and  fragment excitation functions. And the extrapolation ability of the two models will also be tested.

\subsection{ISOTOPIC DISTRIBUTIONS}
\label{isotopic}

\begin{figure}
\centering
\includegraphics[width=8.6cm]{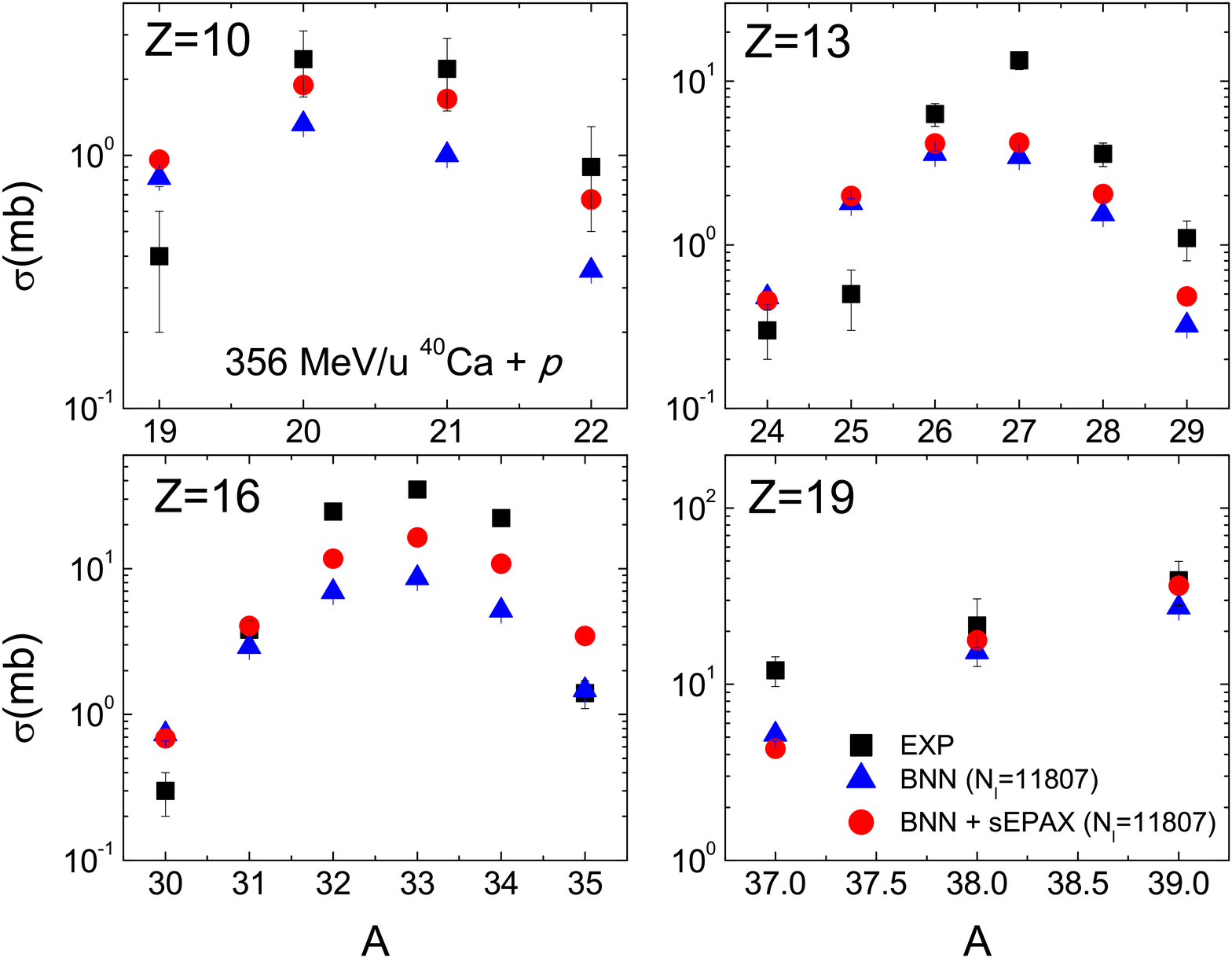}
\caption{(Color online) Test isotopic cross section distributions by the BNN and BNN + sEPAX models with D2 ($N_{l}$=11,807) as the learning set for the 356 MeV/u $^{40}$Ca + $p$ reaction. The experimental and models' error bars are too small to be shown.}
\label{40Ca11807}
\end{figure}

\begin{figure}
\centering
\includegraphics[width=8.6cm]{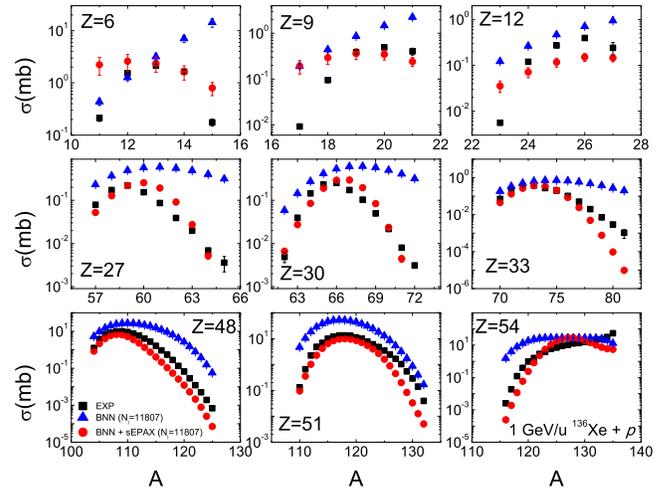}
\caption{(Color online) Similar to FIG. \ref{40Ca11807} but for the 1 GeV/u $^{136}$Xe + $p$ reaction.}
\label{136Xe11807}
\end{figure}

The isotopic distributions for the 356 MeV/u $^{40}$Ca + $p$ and 1 GeV/u $^{136}$Xe + $p$ reactions (see Table \ref{measdata}) are employed to show the performance of BNN and BNN + sEPAX models with $N_{l}$ = 11807. The predicted and measured results are compared in FIG. \ref{40Ca11807} for 356 MeV/u $^{40}$Ca + $p$ reaction and in FIG. \ref{136Xe11807} for 1 GeV/u $^{136}$Xe + $p$ reaction, respectively. For the relatively spallation small system of $^{40}$Ca + $p$ reaction,  the predicted isotopic cross sections by BNN and BNN + sEPAX models are within an order of magnitude difference of the experimental values and their trends of isotopic distributions are consistent. For the $^{136}$Xe + $p$ reaction, which is a relatively large spallation system, the BNN model are poorly reproduce the existing experimental isotopic distributions, because of inadequate fragments in the learning set. While with the physical guidance of sEPAX formula, the predictions by BNN + sEPAX model are in good agreement with experimental data both for the heavy fragments and the quasiprojectile fragments. Poorly predictions for the light neutron-deficient fragments are observed, indicating that further improvement is needed.

\begin{figure}
\centering
\includegraphics[width=8.6cm]{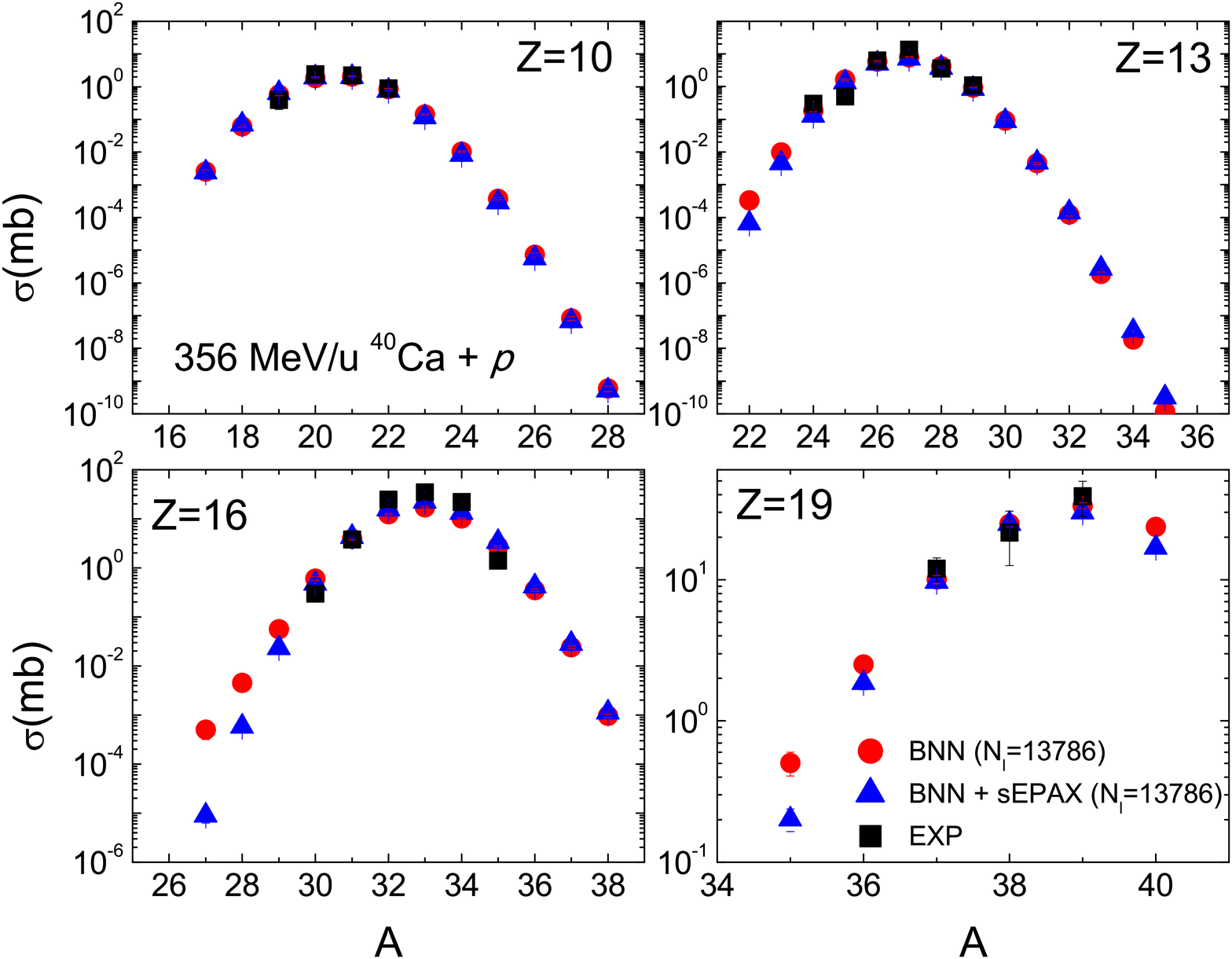}
\caption{(Color online) The extrapolation results of BNN (red circles) and BNN + sEPAX (blue triangles) models with $N_{l}$ = 13,786 learning set for 356 MeV/u $^{40}$Ca + $p$ reaction. The experimental and models' error bars are plotted.}
\label{40Ca13786}
\end{figure}

\begin{figure}
\centering
\includegraphics[width=8.6cm]{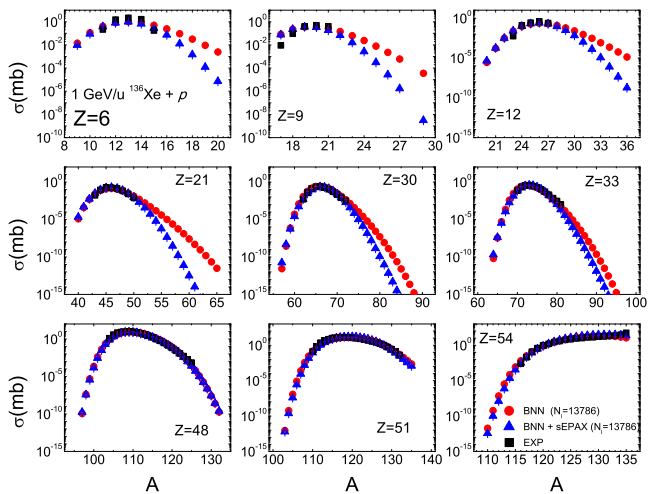}
\caption{(Color online) Similar to Fig. \ref{40Ca11807} but for the 1 GeV/u $^{136}$Xe + $p$ reaction.}
\label{136Xe13786}
\end{figure}

According to the above comparison, it can be concluded that the fragment production mechanisms of dataset D1 and D2 are the same, and the BNN and BNN + sEPAX method can be applied to extrapolate the spallation cross sections. In the following, D1 and D2 are merged (the total number of data is $N_{l}$=13,786) as the new learning set to construct the new BNN and BNN + sEPAX models. In this manner, the weights of for fragments and incident energies in the learning set are improved. The extrapolated isotopic cross sections of the 356 MeV/u $^{40}$Ca + $p$ and 1 GeV/u $^{136}$Xe + $p$ reactions by BNN and  BNN + sEPAX models ($N_{l}$=13,786) are shown in FIGs. \ref{40Ca13786} and \ref{136Xe13786}, respectively.

For the $^{40}$Ca + $p$ reaction, the extrapolations of two models are consistent for the $Z$ = 10 and $Z$ = 13 isotopes, while the BNN model predict much larger cross sections for $Z$ = 16 and $Z$ = 19 neutron-rich isotopes. For the $^{136}$Xe + $p$ reaction, the extrapolations of BNN and BNN + sEPAX models for the neutron-deficient fragments are consistent for the isotopes from $Z=6$ to $Z=54$, while the BNN model predict much larger cross sections than the BNN + sEPAX model for neutron-rich isotopes from the light to medium ones. The extrapolations for the quasiprojectiles are consistent to the measured results. The extrapolation ability of these two models will be further tested later.

\subsection{MASS DISTRIBUTIONS}
\label{massyield}
\begin{figure*}[htbp]
\centering
\subfigure{\includegraphics[width=7cm]{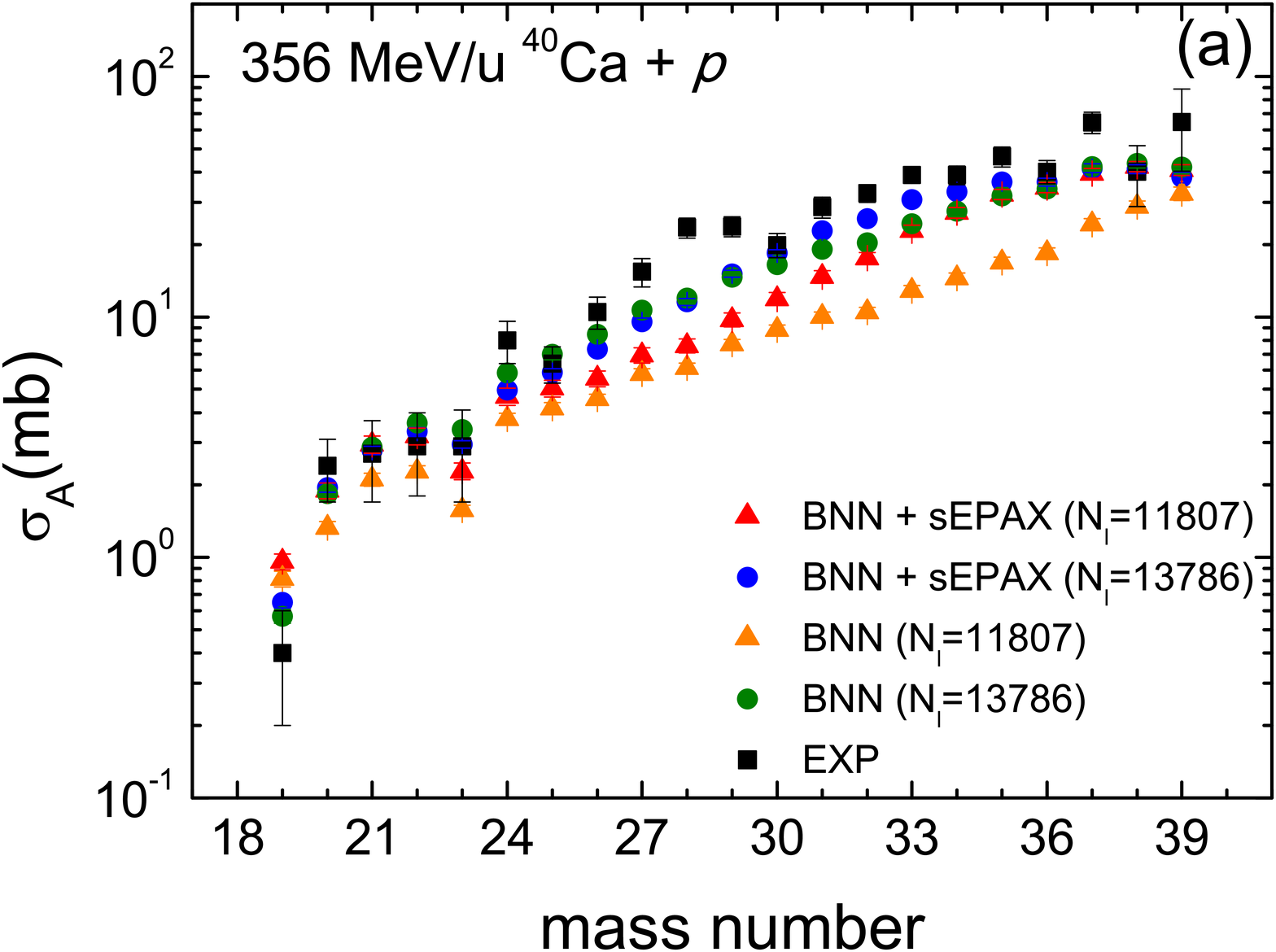}}
\subfigure{\includegraphics[width=7cm]{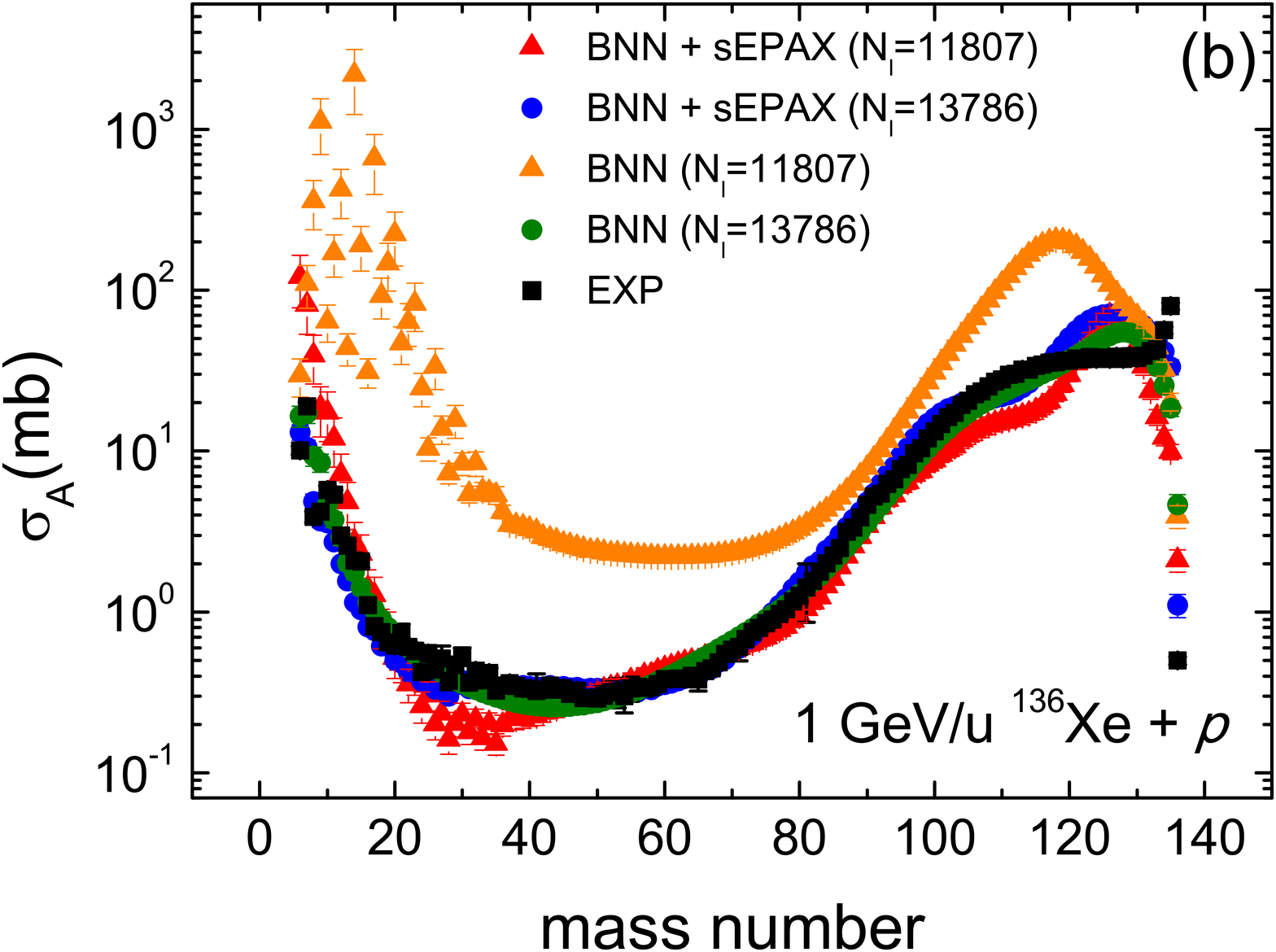}}
\caption{(Color Online) The mass cross sections ($\sigma_{A}$) predicted by BNN and BNN + sEPAX models with different number of learning set ($N_{l} =$ 11,807 and $N_{l} =$ 13,786) for the 356 MeV/u $^{40}$Ca + $p$ [in panel (a)] and 1 GeV/u $^{136}$Xe + $p$ [in panel (b)] reactions. The black squares denote the experimental data. The red triangles (origin triangles) and blue circles (green circles) denote the BNN + sEPAX (BNN) methods with D2 ($N_{l} =$ 11,807) and D1 + D2 ($N_{l} =$ 13,786) as the learning set, respectively.}
\label{mass}
\end{figure*}

The mass yield $Y(A)$ is a key factor for fragment predictions in Eq. (\ref{SigAZ}). It is interesting to see how well the BNN and BNN + sEPAX models can reproduce the mass yield. In FIG. \ref{mass}, the predicted mass distributions for the 356 MeV/u $^{40}$Ca + $p$ and 1 GeV/u $^{136}$Xe + $p$ reactions are shown in panel (a) and (b), respectively. Compared to those models with the $N_{l}$ = 11,807 learning set, the BNN and BNN + sEPAX models using the $N_{l} =$ 13,786 learning set can better reproduce the $\sigma(A)$ distributions for both the two reactions. For the $^{136}$Xe + $p$ reaction, obviously distorted distributions are found in the $A >$ 120 fragments predicted by the BNN and BNN + sEPAX models, for which are very close to the mass number of the projectile nucleus. Since these two reactions are included in the learning set, the result only shows the mass distribution trends of the small and large spallation systems, and cannot demonstrate the generalization ability of the models.

\subsection{Fragment Excitation Functions}
\label{excitaiton}
\begin{figure*}
\centering
\includegraphics[width=15cm]{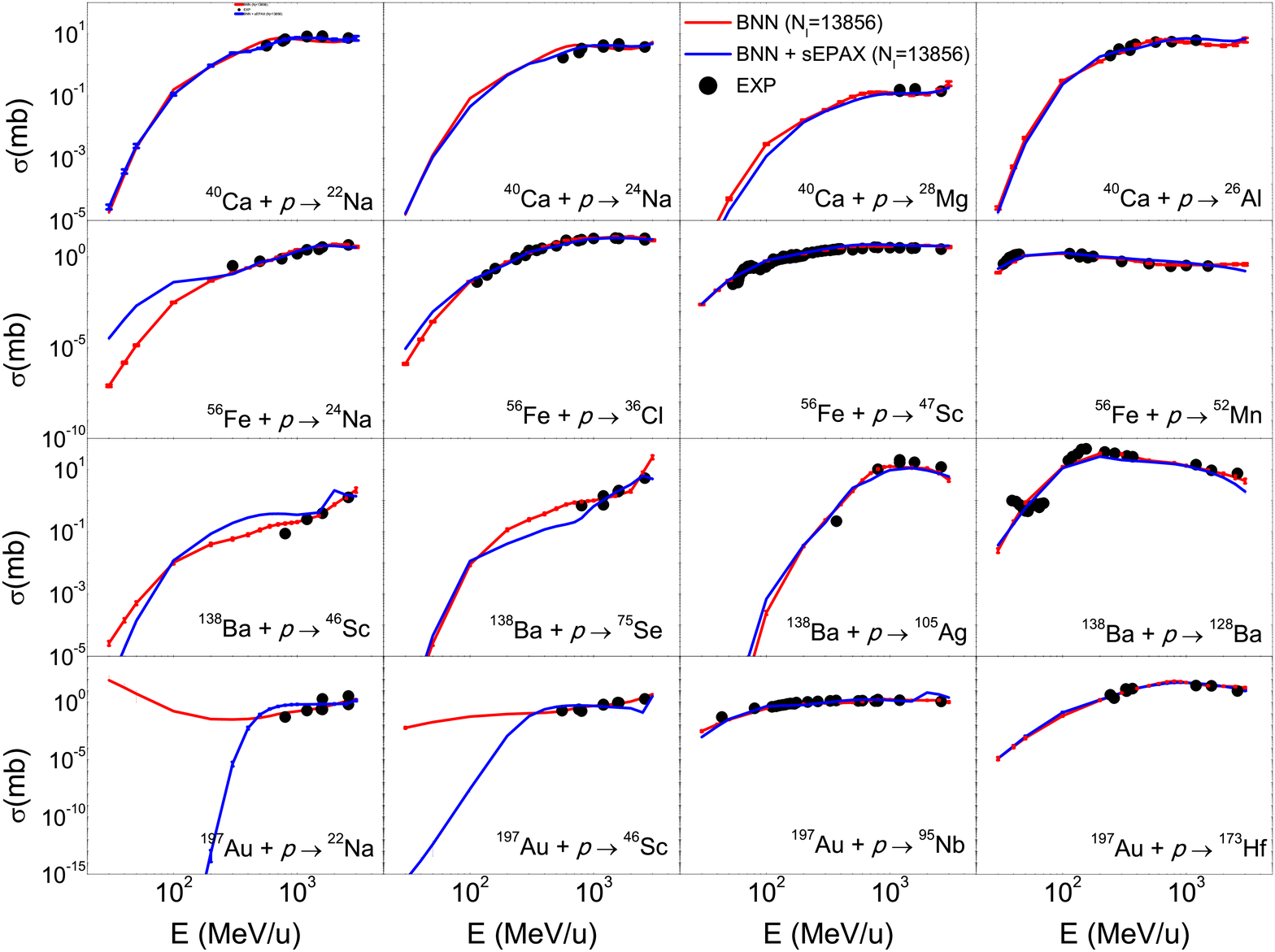}
\caption{(Color online) The fragment excitation functions predicted by BNN and BNN + sEPAX models with $N_{l}$ = 13786 learning set for the $^{40}$Ca + $p$, $^{56}$Fe + $p$, $^{138}$Ba + $p$ and $^{197}$Au + $p$ reactions. The range of incident energy is from 30 MeV/u to 3 GeV/u. The experimental data are taken from Ref. \cite{Michel2014}.}
\label{excurve}
\end{figure*}

It should be noted that, inheriting the main formulas of EPAX, the sEPAX formulas do not incorporate the incident energy term. The incident energy dependence of fragments in the BNN and BNN + sEPAX models are born from the learning of massive data in D2, which makes it possible to yield the excitation functions of fragments in reactions. The excitation functions for the residual fragments have been predicted for the $^{22}$Na, $^{24}$Na, $^{28}$Mg, $^{26}$Al in the $^{40}$Ca + $p$ reactions, $^{24}$Na, $^{36}$Cl, $^{47}$Sc, and $^{52}$Mn in the $^{56}$Fe + $p$ reactions, $^{54}$Mn, $^{75}$Se, $^{105}$Ag, $^{138}$Ba in the $^{138}$Ba + $p$ reactions, and $^{85}$Sr, $^{95}$Nb, $^{160}$Er, $^{173}$Hf in the $^{197}$Ag + $p$ reactions (the data of the measured excitation functions are taken from Ref. \cite{Michel2014}). The range of the incident energy has been selected to be from 30 MeV/u to 3 GeV/u. The selected incident energies range from 30 MeV/u to 3 GeV/u. The results are shown in FIG. \ref{excurve}.
It can be seen that predicted excitation functions well reproduce the experimental data in the four reactions. Considering the mass number of projectile nucleus, for the relatively small system (for example, $^{40}$Ca and $^{56}$Fe), the extrapolation results of BNN and BNN + sEPAX model are consistent, while for the heavy system, relatively large differences emerge in the incident energy below 300 MeV/u (for example, the $^{197}$Ag). Considering the mass number of fragments, the extrapolation of heavy projectile fragments and quasiprojectiles fragments by the BNN and BNN + sEPAX models are consistent with the measured results, but large differences arise for the light and medium fragments (for example,  $^{24}$Na in the $^{56}$Fe reaction, $^{46}$Sc and $^{75}$Se in the $^{138}$Ba reactions). In the $^{197}$Au + $p$ reactions, the BNN and BNN + sEPAX predict different trends for the $^{22}$Na and $^{46}$Sc fragments when the incident energy is below 400 MeV/u. This could be caused by the reason that no data from system larger than $^{136}$Xe is included in the learning set except $^{197}$Au.


\subsection{Model validation}
Due to the large difference between the extrapolations for the BNN and BNN + sEPAX models, an empirical relationship between the cross section and the average binding energy per nucleon of fragment $B'$ in spallation reaction is adopted to test the extrapolation abilities.
\begin{equation}
\sigma = Cexp[(B'-8)/\tau],\\
\end{equation}
where $C$ and $\tau$ are free parameters. $B' = (B - \varepsilon_{p})/A$, in which $\varepsilon_{p}$ is the pairing energy,
\begin{equation}
\varepsilon_{p} = 0.5[(-1)^N + (-1)^Z]\varepsilon_{0} \cdot A^{-3/4}.
\end{equation}

\begin{figure*}
\centering
\includegraphics[width=15cm]{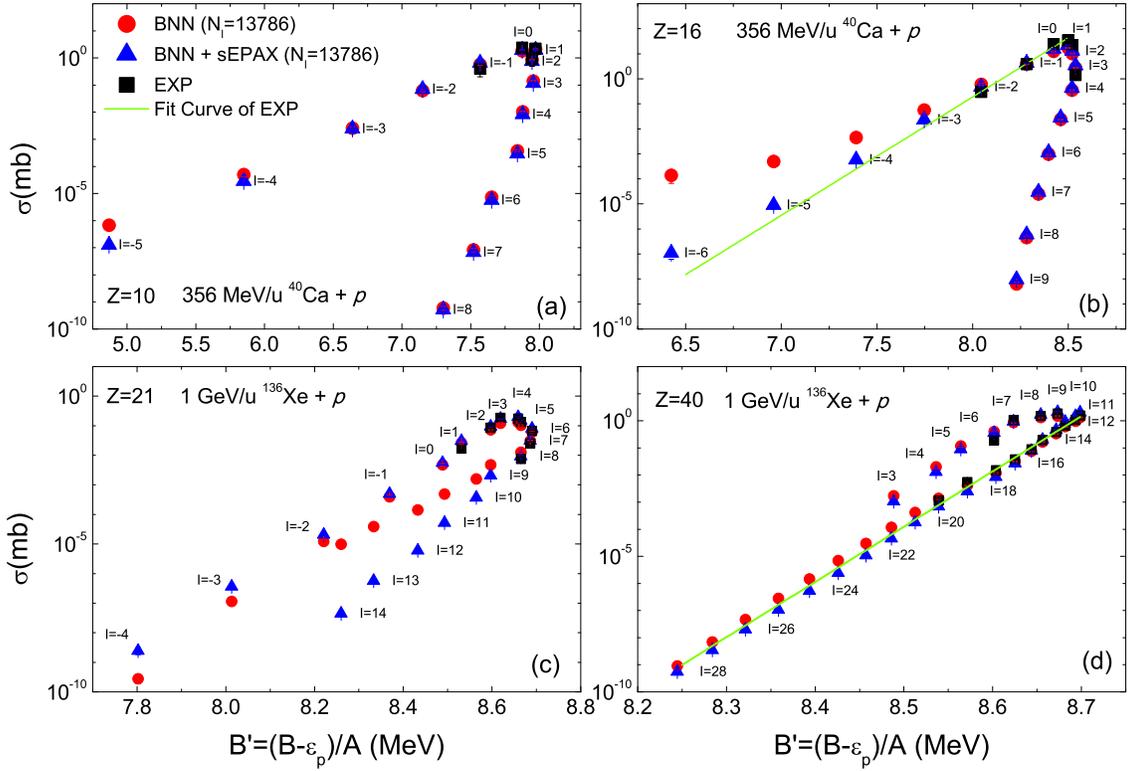}
\caption{(Color online) The correlation between $\sigma$ and B' for isotopes with $Z$ = 10 (in pale (a)), 16 (in pale (b)) in 356 MeV/u$^{40}$ca + $p$ and for $Z$ = 21 (in pale (c)), 40 (in pale (d)) in 1 GeV/u $^{136}$Xe + $p$. The black squares, red circles and blue triangles denote the experimental data, BNN and BNN + sEPAX predictions, respectively. The lines denote the linear fit of the partial experimental points}
\label{Binding}
\end{figure*}

Based on the canonical ensemble theory, this empirical formula has been shown to be reasonable for both neutron-deficient and neutron-rich fragments in multi-fragmentation reaction \cite{Tsang2007,SongYD2018,MaCW2018,MaCW2019,SongYD2019}. The $\sigma\sim B'$ correlation for fragments by BNN and BNN + sEPAX models of $Z =$ 10 and 16 isotopes in 356 MeV/u $^{40}$Ca + $p$ reaction and of $Z =$ 21 and 40 isotopes in 1 GeV/u $^{136}$Xe + $p$ reaction are plotted in Fig. \ref{Binding}. The binding energy of isotopes are taken from AME2020 \cite{AME2020} and $\varepsilon_{0} =$ 30 MeV is adopted \cite{Tsang2007}. Due to the limited experimental data, only the fitting lines for the $Z =$ 16 neutron-deficient isotopes (see panel (b)) and $Z =$ 40 neutron-rich isotopes (see panel (d)) are plotted.

In Fig. \ref{40Ca13786}, it is shown that the extrapolations of BNN and BNN + sEPAX models for the $Z$ = 10 isotopes produced in $^{40}$Ca + $p$ reaction are consistent, and the extrapolations of BNN model are higher than BNN + sEPAX model for $Z =$ 16 neutron-deficient isotopes. In FIG. \ref{Binding}(b), the BNN + sEPAX model predictions agree the fitting function better than the BNN model. In FIG. \ref{136Xe13786}, it is seen that the extrapolations of BNN model are higher than BNN + sEPAX model for $Z <$ 33 neutron-rich isotopes, while both of them are consistent for $Z >$ 33 isotopes. Due to the deficient experimental, the linear fitting results in panel (c) cannot be given, but it is obvious that the BNN predictions for $I >$ 10 fragments (red circles) are excessively upturned. While in panel (d), it is shown that the extrapolations of BNN and BNN + sEPAX models for the $Z =$ 40 neutron-rich isotopes should be both reasonable.

Based on the above discussions, it is suggested that the BNN + sEPAX model with $N_{l}$ = 13,786 learning set can provide precise predictions to fragment cross sections for proton-induced spallation systems smaller than $^{136}$Xe within the incident energy range from tens of MeV/u to above a few GeV/u. Considering the related nuclear applications, it can be used to the residual fragments in proton therapy (above tens of MeV/u to 1 GeV/u), the ADS system and nuclear waste disposal (below 1 GeV/u), solar cosmic ray physics (sub-GeV region and even higher \cite{solarcosmicrayppnp}), shielding at accelerator, etc.
\section{summary}
\label{summary}
The BNN and BNN + sEPAX methods are applied to construct new predictive models for fragment cross sections in proton-induced spallation reactions. Two types of data have been adopted to constructed the predictive models. One type is the isotopic cross sections with incident energy from around 100 MeV/u to 1 GeV/u (D1), and the other type is the fragment excitation functions in reactions of from 30 MeV/u to 2.6 GeV/u (D2). The BNN + sEPAX method
can achieve reasonable extrapolations with less information compared with BNN method. It is suggested that the BNN + sEPAX model based on D1 + D2 learning dataset can be applied to predict the isotopic cross sections and fragment excitation functions  for the proton induced reactions within an incident energy range from tens of MeV/u to a few GeV/u for system smaller than $^{136}$Xe, which can provide precision predictions to both nuclear physics and the related disciplines in nuclear astrophysics, proton therapy, nuclear energy and radioactive ion beam (RNB) facilities.

Furthermore, with plenty of data available, the Bayesian approaches are efficient to provide predictions in various areas. A similar prediction requirement has been emerged in the modern radioactive nuclear beam experiments which search the rare isotopes near/beyond the drip lines using the projectile fragmentation reaction \cite{PF2021PPNP}. For the weak predictive ability in rare isotopes of existing methods, some new semi-empirical methods have been proposed \cite{FRACS,FRACSc,SongYD2019,SCICh192,OESt21MeiPRC}. It is indicative that the empirical formula provided by these works will be very helpful to establish new interesting BNN approaches for projectile fragmentation reactions.

\section*{acknowledgments}
This work is supported by the National Natural Science Foundation of China (Grant No. 11975091, 1210050535 and U1732135), the Program for Innovative Research Team (in Science and Technology) in University of Henan Province (Grant No. 21IRTSTHN011), China.

\end{document}